\documentclass[conference]{IEEEtran}
\IEEEoverridecommandlockouts

\usepackage{cite}
\usepackage{amsmath,amssymb,amsfonts}
\usepackage{algorithmic}
\usepackage{graphicx}
\usepackage{textcomp}
\usepackage{xcolor}
\usepackage{xspace}
\usepackage{booktabs}
\usepackage[hyphens]{url}
\usepackage{hyperref}
\usepackage{comment}
\usepackage{multirow}
\usepackage{multicol}
\usepackage{balance}

\begin{document}
\title{\tool: Harvesting API Information from Various Online Sources}
\newcommand{\tool}{{\textsc{APIHarvest}}\xspace}
\newcommand{\pr}{{pull request}\xspace}


\makeatletter
\newcommand{\linebreakand}{%
  \end{@IEEEauthorhalign}
  \hfill\mbox{}\par
  \mbox{}\hfill\begin{@IEEEauthorhalign}
}
\makeatother

\author{
\IEEEauthorblockN{Ferdian Thung}
\IEEEauthorblockA{\textit{Singapore Management University} \\
Singapore \\
ferdianthung@smu.edu.sg}
\and
\IEEEauthorblockN{Kisub Kim}
\IEEEauthorblockA{\textit{Singapore Management University} \\
Singapore \\
kisubkim@smu.edu.sg}
\and
\IEEEauthorblockN{Ting Zhang}
\IEEEauthorblockA{\textit{Singapore Management University} \\
Singapore \\
tingzhang.2019@phdcs.smu.edu.sg}
\linebreakand
\IEEEauthorblockN{Ivana Clairine Irsan}
\IEEEauthorblockA{\textit{Singapore Management University} \\
Singapore \\
ivanairsan@smu.edu.sg}
\and
\IEEEauthorblockN{Ratnadira Widyasari}
\IEEEauthorblockA{\textit{Singapore Management University} \\
Singapore \\
ratnadiraw.2020@phdcs.smu.edu.sg}
\and
\IEEEauthorblockN{Zhou Yang}
\IEEEauthorblockA{\textit{Singapore Management University} \\
Singapore \\
zyang@smu.edu.sg}
\linebreakand
\IEEEauthorblockN{David Lo}
\IEEEauthorblockA{\textit{Singapore Management University} \\
Singapore \\
davidlo@smu.edu.sg}
}

\maketitle

\pagestyle{plain}

\begin{abstract}
Using APIs to develop software applications is the norm. APIs help developers to build applications faster as they do not need to reinvent the wheel. It is therefore important for developers to understand the APIs that they plan to use. Developers should also make themselves aware of relevant information updates about APIs. In order to do so, developers need to find and keep track of relevant information about the APIs that they are concerned with.  Yet, the API information is scattered across various online sources, which makes it difficult to track by hand. Moreover, identifying content that is related to an API is not trivial. Motivated by these challenges, in this work, we introduce a tool named \tool that aims to ease the process of finding API information from various online sources. \tool is built on works that link APIs or libraries to various online sources. It supports finding API information on GitHub repositories, Stack Overflow's posts, tweets, YouTube videos, and common vulnerability and exposure (CVE) entries; and is extensible to support other sources.

\end{abstract}

\begin{IEEEkeywords}
API mining, API information, multi-source
\end{IEEEkeywords}

\section{Introduction}
APIs are indispensable components of modern software development~\cite{de2004good}. APIs allow developers to focus on developing the main functionality of their software as they can rely on APIs for common functionalities, which consequently lowers software production time and improves developer productivity. Due to its importance, developers should understand the APIs that they plan to use well. They can read API-related information in online sources, such as related code in GitHub, related posts in Stack Overflow, or related tweets in Twitter.

While developers can find the API information themselves, manually finding them could be tedious and time-consuming. First, the API information is scattered across various online sources. Thus, developers would need to open many different websites and perform a search on each one of them. Second, the search functionality in the website is typically not targeted for API search. Thus, developers may need to go through a considerable number of search results before finding the content that is related to the API they search for. Each content may consist of code or text that contains possible mentions of the API they search for. These mentions may be ambiguous, especially if the API name is generic (e.g., \texttt{read}), which may refer to APIs from different libraries or simply an English word \textit{read}. 

Existing works have dealt with the challenges of linking content that is related to either an API or a library while handling the ambiguity that may exist in the possible mentions of API~\cite{luong2022arseek,asyrofi2020ausearch,zhang2022benchmarking,haryono2022automated}.  Luong et al.~\cite{luong2022arseek} proposed an approach that leverages semantic and syntactical analysis to determine whether a Stack Overflow thread is related to an API. Asyrofi et al.~\cite{asyrofi2020ausearch} developed an approach that enables a more accurate code search on GitHub by performing type resolution on the retrieved code and returns only the code related to a particular API. Zhang et al.~\cite{zhang2022benchmarking} investigated the effectiveness of several pre-trained models in the task of determining whether a tweet is related to a library. They found that RoBERTa~\cite{liu2019roberta} performs the best. Haryono et al.~\cite{haryono2022automated} investigated several eXtreme Multi-label Learning (XML) techniques to identify libraries from CVE entries, with LightXML~\cite{jiang2021lightxml} achieving the best performance. 

Although the aforementioned work can be used to identify API content, they can only be used individually. As such, the overall experience is not much of a departure from opening multiple websites at once. Moreover, due to the nature of some contents (e.g., tweets), a direct mention of an API is either extremely rare or non-existent altogether. For such content, it is likely that only a library linking approach is available. In such a case, an extra step of identifying a unique library name that an API belongs to is needed before the available linking approach can be utilized to find API information.

In this work, we built a tool named \tool that can integrate the existing API/library linking approaches in one interface. It aims to streamline the process of finding APIs from many different contents. For content that only has a library linking approach, \tool can automatically map an API to its library in order to leverage the library linking approach. It also supports content that has no linking approach available and defaults to a generic text search. 
With these features, \tool is extensible to support various types of content. 

The rest of the paper is structured as follows: Section~\ref{sec:prelim} discusses the preliminaries that contain some API/library linking approaches that we build our tool upon. Section~\ref{sec:architecture} provides the details of our proposed tool. Section~\ref{sec:impl} describe the evaluation of \tool. Finally, related work and conclusion are presented in Section~\ref{sec:related} and Section~\ref{sec:conclusion}, respectively.

\section{Preliminaries}\label{sec:prelim}
We describe API/library linking approaches that we leverage in \tool. These approaches can be used to identify whether a content is related to an  API or a library.

\subsection{ARSeek}\label{subsec:arseek}
ARSeek~\cite{luong2022arseek} is a method for determining if a post on Stack Overflow is discussing a specific API by analyzing the natural language paragraphs and the code snippets in the post. It has two components: DATYS+ and API Relevance Classifier. The first component, DATYS+, takes in potential posts, which are posts that contain a word matching the simple name of the given API method, and API candidates, which are API methods with the same simple name as the given API method. It then calculates a confidence score for the given API method being referred to in the thread. The second component, API Relevance Classifier, converts natural language paragraphs and code snippets in a Stack Overflow thread; and the comment and implementation code of the given API method into an API Relevance Embeddings. These embeddings are then used to determine the likelihood that the thread is discussing the given API method. A score is then calculated from these embeddings. If the score is above a certain threshold, the thread is considered relevant for the given API method.

\subsection{AUSearch}\label{subsec:ausearch}
Asyrofi et al.~\cite{asyrofi2020ausearch} created AUSearch, an approach that uses type system and API method signatures, which consists of a class name, simple name, and parameters, to improve the precision of GitHub Code Search.\footnote{https://github.com/search} This approach solves the problem of ambiguous API invocations in GitHub code examples, which was a limitation of the GitHub Code Search. By implementing type resolution, AUSearch eliminates irrelevant code examples and significantly increases GitHub Code Search's accuracy for finding code containing the API.

\subsection{Pretrained Models for Library-related Tweets}\label{subsec:librecog}
Zhang et al.~\cite{zhang2022benchmarking} explored several pre-trained models to identify whether a tweet is related to a library. They experimented with both general-purpose (i.e., BERT~\cite{devlin2018bert}, RoBERTa~\cite{liu2019roberta}, and XLNet~\cite{yang2019xlnet}) and domain-specific pre-trained models (i.e., BERTweet~\cite{nguyen2020bertweet} and BERTOverflow~\cite{tabassum2020code}). They fine-tuned these models to the task of classifying library-related tweets. They found that RoBERTa~\cite{liu2019roberta}, which is a robustly optimized BERT~\cite{devlin2018bert}, performs the best and beats existing state-of-the-art.

\subsection{XML Techniques for Identifying Libraries in CVE}\label{subsec:vullinker}
Haryono et al.~\cite{haryono2022automated} explored several eXtreme Multi-label Learning (XML) techniques to identify libraries that are mentioned in CVE entries. They evaluated five traditional models, which are FastXML~\cite{prabhu2014fastxml}, DiSMEC~\cite{babbar2017dismec}, Parabel~\cite{prabhu2018parabel}, Bonsai~\cite{khandagale2020bonsai}, and ExtremeText~\cite{wydmuch2018no}. They also evaluated two deep-learning based models, which are XML-CNN~\cite{liu2017deep} and LightXML~\cite{jiang2021lightxml}. These models are trained on a CVE dataset where the labels are the libraries mentioned across all CVE entries in the dataset. Bonsai and LightXML were shown to be the most effective, with LightXML leading the pack.
\section{Tool Architecture}\label{sec:architecture}
\subsection{Overview}

\begin{figure}[t]
\centerline{\includegraphics[width=\columnwidth]{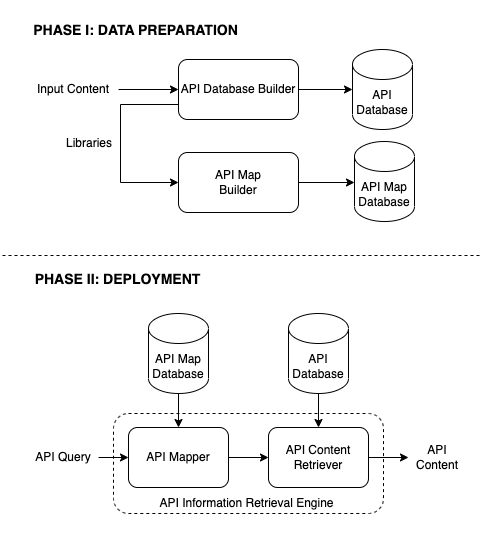}}
\caption{The architecture of \tool}
\label{fig:architecture}
\end{figure}

Figure~\ref{fig:architecture} shows the overall architecture of \tool. \tool consists of three main components: (1) {\em API Database Builder}, (2) {\em API Map Builder}, and (3) {\em API Information Retrieval Engine}. API Database Builder and API Map Builder belong to the {\em Data Preparation} phase, while API Information Retrieval Engine belongs to the {\em  Deployment} phase. 

In the Data Preparation phase, \tool populates the {\em API Database} and {\em API Map Database} with API knowledge from {\em Input Content} and {\em Libraries}, respectively. API Database holds contents from various sources that have been linked to APIs, while API Map Database holds mappings from APIs to their libraries. {\em API Database Builder} uses existing API/library linking approaches (if available) to label the Input Content. It only stores the Input Content to the API Database if the Input Content is labelled as relevant by existing API/library approaches. {\em API Map Builder} adds the mapping from APIs to libraries if only the library linking approach is available to extract information from Input Content. In the Deployment phase, {\em API Information Retrieval Engine} accepts an API query and makes use of the API Database and the API Map Database to return the relevant API content. We further describe \tool components in detail in the following subsections.

\subsection{API Database Builder}\label{sec:preprocess}
Given an Input Content, the API Database Builder first identifies the content source. If there exists an API linking approach that supports the content source, API Database Builder runs the corresponding API linking approach on the Input Content. If the Input Content is labelled as relevant to API(s), the Input Content is stored in the API Database with links to all relevant APIs. 
Similarly, if there exists only a library linking approach that supports the content source, API Database Builder runs the corresponding library linking approach and stores the Input Content to the API Database, with links to all relevant libraries. On the other hand, if there is no API/library linking approach that supports the content source, the Input Content is stored without any links. In all of the above cases, API Database Builder runs a script that adapts the approach's output format to the database format. 
 
\subsection{API Map Builder}\label{sec:update_scripts}
If the Input Content is from a source that is only supported by a library linking approach, API Database Builder sends Libraries that are outputted by the library linking approach to API Map Builder. 
API Map Builder scans all of these libraries and extracts all public APIs from each library (e.g., by using tools such as {\tt javap}\footnote{\url{https://docs.oracle.com/javase/7/docs/technotes/tools/windows/javap.html}}).
It then stores the library name and the public APIs' names in the API Map Database.

\subsection{API Information Retrieval Engine}
Given an API query, API Information Retrieval Engine first passes the query to {\em API Mapper}. The {\em API Mapper} queries the API Map Database for the name of the library that belongs to the API in the query. It then passes both the API name and the library name to the {\em API Content Retriever}. API Content Retriever goes through all the available API information sources (e.g., Stack Overflow, GitHub, etc.) recorded in the API Database. For each source, it queries the API Database using either the API or the library name (depending on the linking approaches used for the source). If the API linking approach was used, it queries the database using the API name. Similarly, if library linking approach was used, it queries using the library name. On the other hand, if neither linking approaches were used, API Content Retriever leverages a classical information retrieval strategy to retrieve relevant contents. It computes and ranks BM25 score~\cite{robertson2009probabilistic} between all contents from the source and the API name. 
After API Content Retriever obtains relevant contents from each source, it aggregates all contents and returns them as the API Content.

\begin{figure}[t]
\centerline{\includegraphics[width=\columnwidth]{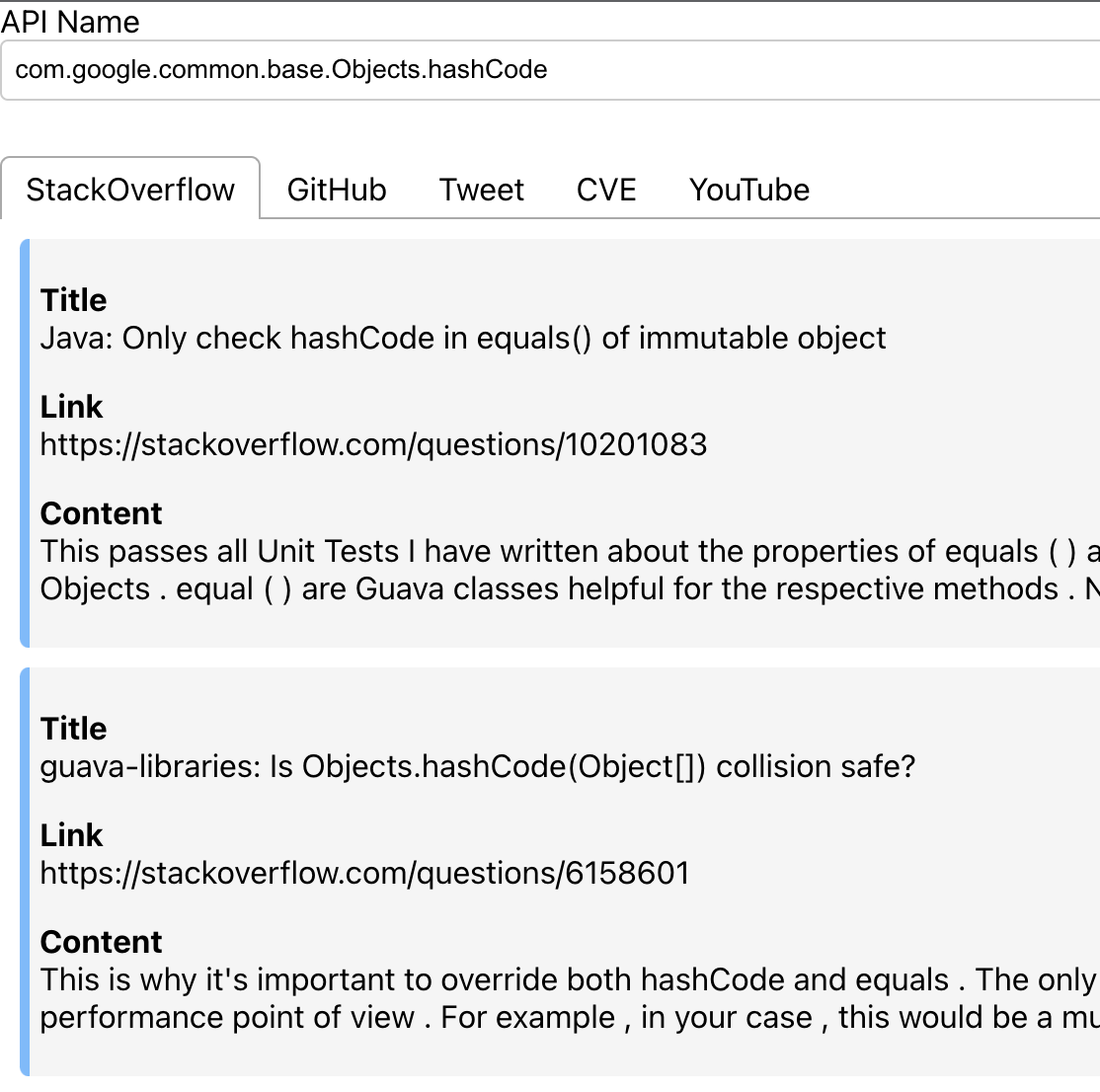}}
\caption{The user interface of \tool}
\label{fig:user_interface}
\end{figure}
Figure~\ref{fig:user_interface} shows the user interface when searching contents related to an API method {\tt com.google.common.base.Object.hashCode()}. The interface allows users to type the API name into the search bar. Once the user ceases typing, the API name is sent to API Information Retrieval Engine, which retrieves the related contents and show them to the user. The contents from different sources are organized in the form of tabs: StackOverflow, GitHub, Tweet, CVE, and YouTube. Each tab contains the relevant API content from each source (if any). The figure shows examples of relevant content in StackOverflow for the given API name.

\section{Evaluation}\label{sec:impl}
\subsection{API Information Sources}
For evaluation and prototype, we integrate the following information sources into \tool: Stack Overflow posts, GitHub code snippets, tweets, CVE entries, and YouTube videos. We run ARSeek~\cite{luong2022arseek} and AUSearch~\cite{asyrofi2020ausearch} to link APIs to Stack Overflow posts and GitHub code snippets, respectively.  Similarly, we run Zhang et al. approach~\cite{zhang2022benchmarking} and Haryono et al. approach~\cite{haryono2022automated} to link software libraries to the tweets and CVE entries, respectively. We use the dataset from their studies and input the test set as the Input Content. We also use the implementation provided in their replication package. For each source, we wrote a script that adapts the data format in the linked content to the database format. Note that we also need to write such a script if we want to extend \tool to support new API/library linking approaches.

For YouTube videos, we randomly selected 100 classes from Java Standard Library.\footnote{\url{https://docs.oracle.com/javase/7/docs/api/overview-summary.html}} We then query YouTube Search API and take the top-100 video metadata returned by the API. We then wrote a script to input these metadata into the API Database. As there is no available API/library linking approaches here, \tool would perform an information retrieval based on the BM25 score.

\subsection{Usability}
We evaluate \tool in the terms of its usability. We do not evaluate the linking effectiveness as it is inherently inherited from the respective API/library linking approaches. We invited 3 Ph.D. students in Computer Science and 2 Research Engineers in Computer Science to evaluate the usability of \tool. All of them have a programming experience of more than 5 years.
We provide them access to \tool and asked them to score three aspects of whether: \tool is easy-to-use, \tool is useful, and they would like to use \tool in the future. The score ranges from 1 to 5, which indicates strongly disagree, disagree, neutral, agree, and strongly agree. The evaluators consider \tool to be easy-to-use (4.8 out of 5) and useful (4.6 out of 5)
and are willing to use \tool in the future (4 out of 5). 

\section{Related Work}\label{sec:related}
\subsection{API Linking Approaches}
API linking approaches aim to link an API to content that is related to it. A group of approaches aims to link an API to a code snippet~\cite{subramanian2014live,saifullah2019learning,phan2018statistical,asyrofi2020ausearch}. To identify API mentions in code snippets, Baker~\cite{subramanian2014live} iteratively performs a deductive linking analysis. Both COSTER~\cite{saifullah2019learning} and STATTYPE~\cite{phan2018statistical} work by capturing and learning the tokens surrounding a code element and associating these tokens with similar tokens that they have encountered before. On the other hand, AUSearch~\cite{asyrofi2020ausearch} leverages type resolution to resolve the API that is being called from within a code.

Another group of approaches aims to link an API to software engineering texts such as Stack Overflow posts~\cite{antoniol2002recovering,marcus2003recovering,bacchelli2010linking,dagenais2012recovering,luong2022arseek}. In this line of work, several studies~\cite{antoniol2002recovering,marcus2003recovering,bacchelli2010linking,dagenais2012recovering} make use of classical information retrieval techniques and/or some heuristics.
Miler, which was developed by Bacchelli et al. ~\cite{bacchelli2010linking}, used string matchings and IR techniques to link emails to source code entities in software systems such as classes in object-oriented systems and functions in procedural language systems.
Dagenais and Robillard~\cite{dagenais2012recovering} used filtering heuristics to link APIs mentioned in software support channels (e.g., mailing lists and forums). ARSeek, which is developed by Luong et al.~\cite{luong2022arseek}, performs a semantic and a syntactical analysis on a Stack Overflow post to determine if the post is related to an API.

\subsection{Library Linking Approaches}
Library linking approaches aim to link a library to content that is related to it. A group of approaches can be used to link a library to microblogs such as tweets~\cite{prasetyo2012automatic, sulistya2020sieve,  zhang2022benchmarking}. Prasetyo et al.~\cite{prasetyo2012automatic} proposed an approach to identify whether a microblog is relevant to software. They extracted features from textual content and URLs in tweets. They then trained a classifier using Support Vector Machine (SVM), which is used to classify whether software-related tweets.  Sulistya et al.~\cite{sulistya2020sieve} proposed an approach to exploit knowledge from rich platforms such as Stack Exchange and Stack Overflow to be used in other platforms such as Twitter. They extracted word embeddings from Stack Exchange and Stack Overflow; and used these embeddings as features in a classifier that identifies software-related tweets. These two approaches can be used to identify library-related tweets by filtering software-related tweets that contain library names. More recently, Zhang et al.~\cite{zhang2022benchmarking} explored the capability of pre-trained models to identify library-related tweets and found them to be more effective than baseline approaches. 

Another group of approaches aims to link a library to CVE entries~\cite{chen2020automated, haryono2022automated}. Chen et al.~\cite{chen2020automated} is the first to propose to automatically link a library that is related to vulnerabilities mentioned in CVE entries. They cast the problem as eXtreme Multi-label Learning (XML), where libraries are the labels, and used FastXML~\cite{prabhu2014fastxml} to identify libraries used in CVE entries. Haryono et al.~\cite{haryono2022automated} continued this direction by evaluating many XML algorithms and found that LightXML~\cite{jiang2021lightxml} is the best XML algorithm for this task. 
\section{Conclusion and Future Work}\label{sec:conclusion}
API information is scattered in various online sources. Linking them to APIs is not a trivial task 
as some mentions of APIs may be ambiguous due to the fact that some APIs have the same method name. Although approaches have been developed to perform the linking, they are typically designed to target only a specific source. For some sources, there is also no available API linking approach. Due to these issues, we present \tool, a tool to find API information from various online sources and present the API information in a single interface. \tool makes use of both API and library linking approaches to find API information. It also accommodates finding API information for all kinds of sources, regardless of the availability of the API/library linking approach. Our preliminary evaluation supports the usability of \tool.
In the future, we plan to add more online sources by leveraging more API/library linking approaches. We release the source code of \tool at \url{https://github.com/soarsmu/APIHarvest}.

\balance

\bibliographystyle{IEEEtran}
\bibliography{main}

\begin{thebibliography}{10}
\providecommand{\url}[1]{#1}
\csname url@samestyle\endcsname
\providecommand{\newblock}{\relax}
\providecommand{\bibinfo}[2]{#2}
\providecommand{\BIBentrySTDinterwordspacing}{\spaceskip=0pt\relax}
\providecommand{\BIBentryALTinterwordstretchfactor}{4}
\providecommand{\BIBentryALTinterwordspacing}{\spaceskip=\fontdimen2\font plus
\BIBentryALTinterwordstretchfactor\fontdimen3\font minus
  \fontdimen4\font\relax}
\providecommand{\BIBforeignlanguage}[2]{{%
\expandafter\ifx\csname l@#1\endcsname\relax
\typeout{** WARNING: IEEEtran.bst: No hyphenation pattern has been}%
\typeout{** loaded for the language `#1'. Using the pattern for}%
\typeout{** the default language instead.}%
\else
\language=\csname l@#1\endcsname
\fi
#2}}
\providecommand{\BIBdecl}{\relax}
\BIBdecl

\bibitem{de2004good}
C.~R. De~Souza, D.~Redmiles, L.-T. Cheng, D.~Millen, and J.~Patterson, ``How a
  good software practice thwarts collaboration: the multiple roles of apis in
  software development,'' in \emph{Proceedings of the 12th ACM SIGSOFT twelfth
  international symposium on Foundations of software engineering}, 2004, pp.
  221--230.

\bibitem{luong2022arseek}
K.~Luong, M.~Hadi, F.~Thung, F.~Fard, and D.~Lo, ``Arseek: identifying api
  resource using code and discussion on stack overflow,'' in \emph{Proceedings
  of the 30th IEEE/ACM International Conference on Program Comprehension},
  2022, pp. 331--342.

\bibitem{asyrofi2020ausearch}
M.~H. Asyrofi, F.~Thung, D.~Lo, and L.~Jiang, ``Ausearch: Accurate api usage
  search in github repositories with type resolution,'' in \emph{2020 IEEE 27th
  International Conference on Software Analysis, Evolution and Reengineering
  (SANER)}.\hskip 1em plus 0.5em minus 0.4em\relax IEEE, 2020, pp. 637--641.

\bibitem{zhang2022benchmarking}
T.~Zhang, D.~P. Chandrasekaran, F.~Thung, and D.~Lo, ``Benchmarking library
  recognition in tweets,'' in \emph{2022 IEEE/ACM 30th International Conference
  on Program Comprehension (ICPC)}, 2022, pp. 343--353.

\bibitem{haryono2022automated}
S.~A. Haryono, H.~J. Kang, A.~Sharma, A.~Sharma, A.~Santosa, A.~M. Yi, and
  D.~Lo, ``Automated identification of libraries from vulnerability data: Can
  we do better?'' 2022.

\bibitem{liu2019roberta}
Y.~Liu, M.~Ott, N.~Goyal, J.~Du, M.~Joshi, D.~Chen, O.~Levy, M.~Lewis,
  L.~Zettlemoyer, and V.~Stoyanov, ``Roberta: A robustly optimized bert
  pretraining approach,'' \emph{arXiv preprint arXiv:1907.11692}, 2019.

\bibitem{jiang2021lightxml}
T.~Jiang, D.~Wang, L.~Sun, H.~Yang, Z.~Zhao, and F.~Zhuang, ``Lightxml:
  Transformer with dynamic negative sampling for high-performance extreme
  multi-label text classification,'' in \emph{Proceedings of the AAAI
  Conference on Artificial Intelligence}, vol.~35, no.~9, 2021, pp. 7987--7994.

\bibitem{devlin2018bert}
J.~Devlin, M.-W. Chang, K.~Lee, and K.~Toutanova, ``Bert: Pre-training of deep
  bidirectional transformers for language understanding,'' \emph{arXiv preprint
  arXiv:1810.04805}, 2018.

\bibitem{yang2019xlnet}
Z.~Yang, Z.~Dai, Y.~Yang, J.~Carbonell, R.~R. Salakhutdinov, and Q.~V. Le,
  ``Xlnet: Generalized autoregressive pretraining for language understanding,''
  \emph{Advances in neural information processing systems}, vol.~32, 2019.

\bibitem{nguyen2020bertweet}
D.~Q. Nguyen, T.~Vu, and A.~T. Nguyen, ``Bertweet: A pre-trained language model
  for english tweets,'' \emph{arXiv preprint arXiv:2005.10200}, 2020.

\bibitem{tabassum2020code}
J.~Tabassum, M.~Maddela, W.~Xu, and A.~Ritter, ``Code and named entity
  recognition in stackoverflow,'' \emph{arXiv preprint arXiv:2005.01634}, 2020.

\bibitem{prabhu2014fastxml}
Y.~Prabhu and M.~Varma, ``Fastxml: A fast, accurate and stable tree-classifier
  for extreme multi-label learning,'' in \emph{Proceedings of the 20th ACM
  SIGKDD international conference on Knowledge discovery and data mining},
  2014, pp. 263--272.

\bibitem{babbar2017dismec}
R.~Babbar and B.~Sch{\"o}lkopf, ``Dismec: Distributed sparse machines for
  extreme multi-label classification,'' in \emph{Proceedings of the tenth ACM
  international conference on web search and data mining}, 2017, pp. 721--729.

\bibitem{prabhu2018parabel}
Y.~Prabhu, A.~Kag, S.~Harsola, R.~Agrawal, and M.~Varma, ``Parabel: Partitioned
  label trees for extreme classification with application to dynamic search
  advertising,'' in \emph{Proceedings of the 2018 World Wide Web Conference},
  2018, pp. 993--1002.

\bibitem{khandagale2020bonsai}
S.~Khandagale, H.~Xiao, and R.~Babbar, ``Bonsai: diverse and shallow trees for
  extreme multi-label classification,'' \emph{Machine Learning}, vol. 109,
  no.~11, pp. 2099--2119, 2020.

\bibitem{wydmuch2018no}
M.~Wydmuch, K.~Jasinska, M.~Kuznetsov, R.~Busa-Fekete, and K.~Dembczynski, ``A
  no-regret generalization of hierarchical softmax to extreme multi-label
  classification,'' \emph{Advances in neural information processing systems},
  vol.~31, 2018.

\bibitem{liu2017deep}
J.~Liu, W.-C. Chang, Y.~Wu, and Y.~Yang, ``Deep learning for extreme
  multi-label text classification,'' in \emph{Proceedings of the 40th
  international ACM SIGIR conference on research and development in information
  retrieval}, 2017, pp. 115--124.

\bibitem{robertson2009probabilistic}
S.~Robertson, H.~Zaragoza \emph{et~al.}, ``The probabilistic relevance
  framework: Bm25 and beyond,'' \emph{Foundations and Trends{\textregistered}
  in Information Retrieval}, vol.~3, no.~4, pp. 333--389, 2009.

\bibitem{subramanian2014live}
S.~Subramanian, L.~Inozemtseva, and R.~Holmes, ``Live api documentation,'' in
  \emph{ICSE}, 2014.

\bibitem{saifullah2019learning}
C.~K. Saifullah, M.~Asaduzzaman, and C.~K. Roy, ``Learning from examples to
  find fully qualified names of api elements in code snippets,'' in
  \emph{ASE}.\hskip 1em plus 0.5em minus 0.4em\relax IEEE, 2019.

\bibitem{phan2018statistical}
H.~Phan, H.~A. Nguyen, N.~M. Tran, L.~H. Truong, A.~T. Nguyen, and T.~N.
  Nguyen, ``Statistical learning of api fully qualified names in code snippets
  of online forums,'' in \emph{ICSE}.\hskip 1em plus 0.5em minus 0.4em\relax
  IEEE, 2018.

\bibitem{antoniol2002recovering}
G.~Antoniol, G.~Canfora, G.~Casazza, A.~De~Lucia, and E.~Merlo, ``Recovering
  traceability links between code and documentation,'' \emph{TSE}, vol.~28,
  no.~10, 2002.

\bibitem{marcus2003recovering}
A.~Marcus and J.~I. Maletic, ``Recovering documentation-to-source-code
  traceability links using latent semantic indexing,'' in \emph{ICSE}.\hskip
  1em plus 0.5em minus 0.4em\relax IEEE, 2003.

\bibitem{bacchelli2010linking}
A.~Bacchelli, M.~Lanza, and R.~Robbes, ``Linking e-mails and source code
  artifacts,'' in \emph{ICSE}, 2010.

\bibitem{dagenais2012recovering}
B.~Dagenais and M.~P. Robillard, ``Recovering traceability links between an api
  and its learning resources,'' in \emph{ICSE}.\hskip 1em plus 0.5em minus
  0.4em\relax IEEE, 2012.

\bibitem{prasetyo2012automatic}
P.~K. Prasetyo, D.~Lo, P.~Achananuparp, Y.~Tian, and E.-P. Lim, ``Automatic
  classification of software related microblogs,'' in \emph{2012 28th IEEE
  International Conference on Software Maintenance (ICSM)}.\hskip 1em plus
  0.5em minus 0.4em\relax IEEE, 2012, pp. 596--599.

\bibitem{sulistya2020sieve}
A.~Sulistya, G.~A.~A. Prana, A.~Sharma, D.~Lo, and C.~Treude, ``Sieve: Helping
  developers sift wheat from chaff via cross-platform analysis,''
  \emph{Empirical Software Engineering}, vol.~25, no.~1, pp. 996--1030, 2020.

\bibitem{chen2020automated}
Y.~Chen, A.~E. Santosa, A.~Sharma, and D.~Lo, ``Automated identification of
  libraries from vulnerability data,'' in \emph{Proceedings of the ACM/IEEE
  42nd International Conference on Software Engineering: Software Engineering
  in Practice}, 2020, pp. 90--99.

\end{thebibliography}

\end{document}